\documentstyle[12pt,epsf]{article}

\begin{document}
\def\beq{\begin{equation}}
\def\eeq{\end{equation}}
\def\beqa{\begin{eqnarray}}
\def\eeqa{\end{eqnarray}}
\def\noin{\noindent}
\def\grad{{\bf \nabla}}
\def\bv{{\bf v}}
\def\bB{{\bf B}}
\def\bJ{{\bf J}}
\def\bE{{\bf E}}
\def\bK{{\bf K}}
\def\bx{{\bf x}}
\def\pa{\partial}
\def\eps{\epsilon_{\alpha\beta}}
\def\rta{\rightarrow}
\def\lra{\leftrightarrow}
\def\tm{\tilde{m}}
\def\tn{\tilde{n}}
\def\tbv{\tilde{\bv}}
\def\tbK{\tilde{\bK}}
\def\tP{\tilde{P}}
\def\trho{\tilde{\rho}}
\def\hm{\hat{m}}
\def\hn{\hat{n}}
\def\hbv{\hat{\bv}}
\def\hbK{\hat{\bK}}
\def\hP{\hat{P}}
\def\hrho{\hat{\rho}}
\def\cE{{\bf {\cal E}}}
\def\cB{{\bf {\cal B}}}
\def\bbJ{\bar{\bJ}}
\def\brho{\bar{\rho}}
\def\bsig{\bar{\sigma}}
\def\ccB{{\cal B}}
\def\ccE{{\cal E}}
\def\ccA{{\cal A}}
\def\hcA{\hat{\ccA}}
\titlepage
\begin{flushright} QMW-PH-96-05
\end{flushright}
\vspace{3ex}
\begin{center} \bf
{\bf CONFORMAL  SOLUTIONS OF DUALITY INVARIANT
2D MAGNETOHYDRODYNAMIC TURBULENCE}
\\
\rm
\vspace{3ex}
Omduth Coceal$^{a}$\footnote{e-mail: o.coceal@qmw.ac.uk},
Wafic A. Sabra$^{b}$\footnote{e-mail:
uhap012@vax.rhbnc.ac.uk} and
Steven Thomas$^{a}$\footnote{e-mail: s.thomas@qmw.ac.uk}\\
\vspace{2ex}

$^{a}${\it Department of Physics\\
Queen Mary and Westfield College\\
Mile End Road\\
London E1 4NS\\
U.K.}\\
\vspace{4ex}

$^{b}${\it Department of Physics\\
Royal Holloway and Bedford New College\\
Egham\\
Surrey TW20 0EX\\
U.K.}\\

\vspace{4ex}
ABSTRACT
\end{center}
\noindent

We consider possible conformal field theory (CFT) descriptions of
the various inertial ranges that exist in $2d$ duality invariant
Magnetohydrodynamics. Such models arise as  effective theories of
dyonic plasmas in 3 dimensions in which all fields are independent
of the third coordinate. We find new constraints on the allowed CFT's
compared to those that may describe turbulence in  $2d$ plasmas of
electric charges only. The predictions from CFT concerning
equipartition of energy  amongst the electric and magnetic fields are
discussed, and quantities exhibiting universal scaling are derived.
\newpage
Recently, we derived a generalized  theory of Magnetohydodynamics by
incorporating, in a variety of ways, magnetic charges [1].
The duality invariant Maxwell's equations (see [4] for example) together with
fluid dynamics are the main ingredients in this generalization. The plasma
physics of purely magnetic charges can also be obtained by performing  duality
transformations on all the equations of
ordinary MHD.\footnote {For some recent applications of dual MHD see [6]}
 Moreover, it was also shown in [1] that a mixed plasma of both electric and
magnetic charges, in the ideal limit, reduces to a fluid with vanishing
electric and magnetic fields, while a plasma of dyons
(which carry both electric and magnetic charge) exhibit new physical
properties. Motivated by recent work of Polyakov [7], we would like to
understand the role of CFT in finding solutions to the above mentioned cases in
the ideal limit.

The analysis of the magnetic plasma case, being dual to that of electric one,
is the same as in [2,5]  with the role of electric and magnetic field
interchanged. Also, the case of a mixed plasma, being a fluid in the ideal
limit, is covered by the analysis of [7]. Therefore, in this letter, we will
concentrate on the case of dyonic plasmas. For a detailed analysis of the
equations of dyonic MHD we refer the reader to [1].

For the purposes of this paper, we need only to display the equations of dyonic
MHD in the ideal limit. These are given by
\beqa\label{eq:4.23}
\grad\cdot\bv & = & 0, \qquad\grad\cdot\cB  =  0,\qquad  \cE =
\frac{\displaystyle 1}{\displaystyle c} \bv \times \cB,\cr
\grad\times (\bv\times\cB) & = & \frac{\pa\cB}{\pa t}, \qquad
\grad\times(\frac{\pa\bv}{\pa t}+\bv\cdot\grad \bv)  =
\frac{1}{4\pi\rho_{M}(e^{2}+g^{2})}
\times (\cB\cdot\grad \cB).
\eeqa
The quantities $\cE $ and $\cB$ are duality invariant combinations of the
physical electric and magnetic fields $\bE$ and $\bB$,
defined as
\beqa\label{eq:4.18}
\cE = e\bE+g\bB, \qquad\cB  =  g\bE-e\bB,
\eeqa
here $e$ and $g$ are, respectively,  the electric and magnetic charges, $\bv$
is the velocity and $\rho_M$ is the mass density.
The crucial observation is that, up to factors of $(e^2+g^2)$, these are
exactly the equations of ordinary MHD with suitable field identification.
This implies that the CFT analysis of duality invariant generalized MHD
turbulence follows from that of ordinary MHD [2,5]. However,  we should note
that the duality-invariant fields $\cE$ and $\cB$ are not the physical fields
but rather a linear combination of them.This, as we shall see, in general will
put extra constraints on the allowed  CFT solutions, in order to have a
consistent interpretation of the physical fields in terms of CFT operators.  In
addition, the consequences of duality symmetry
concerning the equipartition of a fraction of the plasmas  energy into the
physical $\bE$ and $\bB$ fields are explored.
For completeness and by way of introducing necessary formulae,
a brief summary of some of the results of [2] is given, but expressed in the
present context.

In order to investigate the possible role of $2d$ CFT in understanding
generalized MHD turbulence, it is natural to consider the latter theory
in two dimensions. An effective two-dimensional theory is defined by
 imposing the
condition that all fields are independent of the third coordinate, $i.e$,
$\pa_{3}\bv=\pa_{3}\cB=0$. The generalized theory of MHD is obtained by
defining  the two-dimensional \lq\lq passive" scalars $v_{3} \equiv V$ and
 $\ccB_{3} \equiv \ccB$, and
the two-dimensional vectors $v_{\alpha}$ and $\ccB_{\alpha}$, with
$\alpha , \beta =1,2$. The latter can be expressed in terms of the
two-dimensional potentials $\psi$ and $\ccA$,
\beq\label{eq:5.2}
v_{\alpha}=\eps\pa_{\beta}\psi,~~~\ccB_{\alpha}=\eps\pa_{\beta}\ccA.
\eeq

In terms of these quantities and $\hcA\equiv \eps\pa_{\beta}\ccA\pa_{\alpha},$
the dimensionally-reduced set of dyonic  MHD equations,  take the form
\beqa\label{eq:5.4}
\dot{\omega} + \eps\pa_{\beta}\psi\pa_{\alpha}\omega & = &
-\frac{1}{4\pi\rho_{M}}\hcA\Box\ccA, \qquad
\dot{\ccA} + \eps\pa_{\beta}\psi\pa_{\alpha}\ccA  =  0, \cr
\dot{V} + \eps\pa_{\beta}\psi\pa_{\alpha}V & = & \frac{1}{4\pi\rho_{M}}\hcA
\ccB, \qquad
\dot{ \ccB} + \eps\pa_{\beta}\psi\pa_{\alpha} \ccB  =  \hcA V,
\eeqa
where $\Box\equiv\pa_{\alpha}\pa_{\alpha},$ the dot over a symbol denotes time
differentiation and
$\omega \equiv \eps \pa_{\alpha} v_{\beta} $ is the $2d$ vorticity .

Polyakov's proposal, in the case of inviscid fluid flow,
was to identify the statistical fields of turbulent fluids with certain
operators
of a CFT such that the  Hopf equations were satisfied, thus offering an exact
solution
to these equations. It should however be emphasized
that such solutions describe a kind of static limit of turbulence where
all correlators involving $t$-derivatives of fields vanish [7].
The Hopf equations resulting from  (\ref{eq:5.4}) are considerably
more complicated than those considered by Polyakov [7], and first attempt at
finding their
CFT solutions was due to Ferreti and Yang [5] and later in [10] where
simplified solutions to pure $2d$ MHD in which  $V = \ccB = 0$ (in the present
notation) were found.  Another set of reduced solutions were obtained in [2]
where the so
called perpendicular flow was considered in which the $3d$ vectors $\bv $ and
$\cB$ are at right angles to each other. One particular realization of this is
to set $\ccA=V=0,$
which gives Hopf equations corresponding to
\beqa\label{eq:5.5}
\dot{\omega}+\eps(\pa_{\beta}\psi)(\pa_{\alpha}\omega)=0, \qquad
\dot{\ccB}+\eps(\pa_{\beta}\psi)(\pa_{\alpha} {\ccB}) =0.
\eeqa
These equations imply the existence of three quadratic conserved quantities:
enstrophy $1/2 \int \omega^{2}~d^{2}x$, kinetic energy
$1/2 \int (v_{\alpha}v_{\alpha})~d^{2}x$ and the quantity
$1/2 \int  {\ccB}^{2}~d^{2}x$.

In the pure $2d$ MHD case mentioned above, the presence of a non-vanishing $2d$
vector $\cB_{\alpha} $ does not allow for a conserved flux of enstrophy
(even if one allows oneself the freedom to add terms in $\cB_{\alpha} $ ), but
there
is a conserved energy flux, $1/2 \int (v_{\alpha}v_{\alpha} +
\frac{\displaystyle 1}{\displaystyle {4 \pi }} \cB_{\alpha}
\cB_{\alpha} )~d^{2}x.$

Before moving on to consider the specific constraints on CFT solutions to the
Hopf equations,
we discuss the additional constraint that arise in satisfying Hopf equations
associated with dyonic plasmas. These come from the fact that  assigning the
appropriate conformal primary or secondary fields to  $v_{\alpha}, V ,
\cB_{\alpha} $
and $\ccB $ to satisfy such equations is not the end of the story. Such an
assignment should be supplemented with further constraints on the operator
product expansion (OPE) structure of CFT such that
physical fields  have a
well defined interpretation.  Moreover,  one of the
predictions of the CFT approach is to obtain the slope of the spectrum of
energy fluctuations
which is related to the conformal dimensions of the primary fields [7].  In the
present
situation we should be able to determine the various contributions
to this spectrum from the physical electric and magnetic fields. Within the CFT
approach,
this simply require an unambiguous assignment of a particular primary field (or
its secondary) to such physical fields for any given solution.

If we express the physical fields $E _{\alpha} , E $ and $B_{\alpha} , B$  in
terms of their duality
invariant counterparts,  one has:
\beqa\label{eq:5.5a}
 E_{\alpha} &=&\lambda\Big( \frac{{\epsilon}_{\alpha \beta} }{cg}
(v_{\beta} \cB - {\cB}_{\beta} V )+
 \frac{{\cB}_{\alpha}}{e} \Big), \quad
E= \lambda\Big( \frac{{\epsilon}_{\alpha \beta} }{cg}
(\partial_{\alpha} \psi \partial_{\beta }\ccA ) +
 \frac{{\cB}}{e}\Big), \cr
B_{\alpha} & = &\lambda\Big( \frac{{\epsilon}_{\alpha \beta }}{ce}
(v_{\beta} \cB - {\cB}_{\beta} V ) -
 \frac{{\cB}_{\alpha}}{g}\Big), \quad
 B  = \lambda\Big( \frac{{\epsilon}_{\alpha \beta }}{ce}
(\partial_{\alpha} \psi \partial_{\beta }\ccA) -\frac{{\cB}}{g}\Big).
\eeqa
where $\lambda={(\frac{\textstyle e}{\textstyle g} + \frac{\textstyle
g}{\textstyle e} {)}^{-1}}.$
Now recall that the fields $\cB_{\alpha} , \cB , v_{\alpha} , V$ will
be associated with specific primary or secondary fields. However,
it is clear from (\ref{eq:5.5a}) that there is in general an ambiguity
in determining the corresponding fields associated to the
physical electric and magnetic fields. In fact if all terms on the rhs of
(\ref{eq:5.5a}) are non-vanishing,  we then need
to place extra conditions on certain
OPE's in the CFT. Specifically if $\psi , \ccA , \cB $ and $ V $ are
identified with primary fields then we require
\beqa\label{eq:5.81}
[\psi \times \cB ] = [ \ccA \times V ] =  [\ccA ],  \qquad
[\psi \times \ccA ]   =  [ \cB ],
\eeqa
where [$\phi $] in (\ref{eq:5.81}) denotes the conformal family of  the primary
field $\phi .$ However, under certain circumstances these additional
constraints do not arise.  For instance, consider the two cases discussed
earlier. First if
$V = \cB = 0 $, then clearly there is no longer any ambiguity because
all quadratic terms in $ E_{\alpha} , B_{\alpha} $ vanish, likewise the linear
terms in $E, B $. In the second limit, $\ccA = V = 0 $,
we also have no ambiguity since this time the linear terms in
 $E_{\alpha} , B_{\alpha} $ vanish, likewise the quadratic terms in
$E, B $. But in general, we should impose further constraints on the
OPE's as in (\ref{eq:5.81}) to make sense of the CFT approach.
Later on in this section we shall see that making unambiguous connections
between
the physical fields and primary fields, allows one to  make predictions
concerning the equipartition of energy  amongst the  fields
$E_{\alpha} , B_{\alpha} $ and the passive scalars $E, B $.

\noin
Let us now return to the specific problem of finding CFT solutions to
our dyonic plasma coupled to passive scalars in $d=2$.
In keeping with Polyakov's original idea, we would like to interpret the
fields $\psi$ and $ \ccB$ as primary operators in some CFT. The vorticity
$\omega\equiv -\pa^{2}\psi$ then becomes a level one field in the conformal
family of $\psi$.
The equations of motion (\ref{eq:5.5}) can be considered as defining equations
for new
fields $\dot{\omega}$ and $\dot{\ccB} $, provided that due account is taken of
short distance effects by implementing the point-split regularization scheme
when a
product of two fields defined at the same point arises. The result is
\beqa\label{eq:5.6}
\dot{\omega}  \sim  |a|^{2(\Delta_{\phi}-2\Delta_{\psi})}
[L_{-2}\bar{L}^{2}_{-1}-\bar{L}_{-2}L^{2}_{-1}]\phi, \quad
\dot{\ccB}   \sim   |a|^{2(\Delta_{\chi}-\Delta_{\ccB}-\Delta_{\psi}+1)}
 [L_{-2}\bar{L}^{2}_{-1}-\bar{L}_{-2}L^{2}_{-1}]\chi,
\eeqa
where $\phi$ and $\chi$ are the minimal dimension operators in the OPE
$\psi\times\psi$ and $\psi\times \ccB$ respectively, the $L_{-n}$'s are the
usual Virasoro generators and $a$ is an UV cutoff.
These relations fix the dimension of $\dot{\omega}$ and $\dot{\ccB}$ to
$2+\Delta_{\psi}$ and $2+\Delta_{\chi}$ respectively.

\noin
As described in greater detail in [2],
we would like to make connection with the theory of turbulence by postulating
that Hopf equations are satisfied by $\omega$ and $ \ccB$.
These imply the vanishing of $\dot{\omega}$ and $\dot{\ccB}$, which leads to
the
inequalities
\beqa\label{eq:5.7}
\Delta_{\phi}  >  2\Delta_{\psi}, \qquad
\Delta_{\chi}  >  \Delta_{\psi}+\Delta_{ \ccB}-1.
\eeqa
The additional physical principle necessary in the description of turbulent
solutions is the requirement that the flux of energy or enstrophy be constant
on infrared scales, consistent with either an energy cascade or enstrophy
cascade scenario.
As was demonstrated by Polyakov [7], constant enstrophy flux implies that
$\Delta_{\psi}+\Delta_{\phi}=-3$, while the corresponding constraint for
constant energy  flux derived by Lowe [8] is $\Delta_{\psi}+\Delta_{\phi}=-2$.
Assuming a constant $ \ccB$-flux, which is consistent with the
conservation of $ \int {\ccB}^{2}~d^{2}x$, an additional constraint
$\Delta_{\ccB}+\Delta_{\chi}=-2$ on the dimensions of $\ccB$ and $\chi$ is
derived.

One can impose the condition of constant $\ccB$-flux
together with either constant enstrophy flux
or constant energy flux. We found only one solution in the former case among
minimal models $(p,q)$ [11] with $q < 500$ and two solutions for constant
energy in the same range. The reader is referred to  [2] for details.

\noin
We now turn to the situation where we consider different limiting
cases of (\ref{eq:5.4}). We have
investigated all possibilities and we found that in nearly all the cases we
obtain
equivalent results to either the case outlined above or that described by
Ferretti and Yang [5], namely the limit $\ccB=V=0$.
The notable exception is if we set $\ccA=0$ without any constraint on $V$. We
then obtain the following equations of motion:
\beqa\label{eq:5.11}
\dot{\omega}+\eps(\pa_{\beta}\psi)(\pa_{\alpha}\omega)= 0, \qquad
\dot{V}+\eps(\pa_{\beta}\psi)(\pa_{\alpha}V) =0, \qquad
\dot{\ccB}+\eps(\pa_{\beta}\psi)(\pa_{\alpha}\ccB) = 0.
\eeqa
We note that the kinetic energy corresponding to the third component of
velocity $V$ is independently conserved.
Hence, in addition to the flux constraints above we obtain an additional
condition which must be satisfied simultaneously, namely,
$\Delta_{V}+\Delta_{\varphi}=-2,$
where $\varphi$ is the minimal field in the OPE of $\psi\times V$.Imposing this
additional constraint makes it even more difficult to find exact solutions for
this system. We have not found any explicit examples
in our searches through non unitary minimal models so far.

\noin
We now consider the consequences  of CFT solutions to the Hopf equations,
along with the flux matching conditions, with regard to the partitioning
of kinetic energy amongst the various fields.  For simplicity we consider
the previously discussed limiting cases  of $ \ccB = V = 0 $, or
$V=\ccA = 0.$ Defining the momentum space energy density
${\cal K}(k)$
\beqa\label{eq:fish}
 \int d k  {\cal K}{(k)} \, & = &  \, \frac{\rho_M}{2} ( <v_{\alpha}
v_{\alpha}>
+ <V V> ) + \frac{1}{4 {\pi}^2 (e^2 + g^2 ) } ( g^2 <E_{\alpha} E_{\alpha}>
\cr
&+& g^2  < E E > + e^2 < B_{\alpha} B_{\alpha} > +
e^2 < B B > \cr
& - & 2 e g ( < B_{\alpha} E_{\alpha} > + <BE>).
\eeqa
For the case $\ccB = V = 0 $, $E_{\alpha} = - \frac{\textstyle e}{\textstyle g}
B_{\alpha} = \frac{\textstyle\lambda}{\textstyle e}{\ccB_{\alpha}}$
and one may show that
\beqa\label{eq:zoo}
\int d k  {\cal K}{(k)} \, & = &  \, \frac{\rho_M}{2} ( < v_{\alpha} v_{\alpha}
>) +
 \frac{1}{4 {\pi}^2  } ( <B_{\alpha} B_{\alpha} > + < E_{\alpha} E_{\alpha} >),
\eeqa
from which we learn that
\beqa\label{eq:no}
{\cal K}{(k)}  \, & = & \,{{\cal K}^{kin}}{(k)} + {{\cal K}^{elec}}{(k)} +
{{\cal K}^{mag}}{(k)},
\cr
{{\cal K}^{kin}}{(k)} & \sim  & k^{1 + 4 \Delta_{\psi} } , \,\,
{{\cal K}^{elec}}{(k)}  \sim k^{1 + 4 \Delta_{\ccA} },\,\,
{{\cal K}^{mag}}{(k)}  \sim k^{1+ 4 \Delta_{\ccA} }
\eeqa
where ${{\cal K}^{kin}}{(k)} $, ${{\cal K}^{elec}}{(k)}$ and
${{\cal K}^{mag}}{(k)}$ are contributions to the energy spectrum
from the kinetic energy $<v_{\alpha} v_{\alpha}>$ and
electric and magnetic energies $< E_{\alpha} E_{\alpha} > ,
< B_{\alpha} B_{\alpha} >$ respectively.
What is apparent from (\ref{eq:no}) is that whilst there does not need to
be an equal partitioning of energy between velocity and electric (or magnetic)
components ($\Delta_{\psi} \neq \Delta_{\ccA} $ in general), there is an
equipartition of a fraction of the total kinetic energy amongst the
electric and magnetic fields. The latter is clearly a consequence of
the existence of electric-magnetic duality.

Finally we discuss the generalization to dyonic MHD of the universal scaling
invariants first discussed in [9] in the context of Polyakov turbulence.
There it was shown that irrespective of any particular CFT solution to
turbulence,
the quantity
\beq\label{eq: 600}
Q(k) =( E(k) P(k) k ^{-8} {)}^{\frac{1}{4} } \sim k^{-3}
\eeq
is a universal scaling invariant,
where $\int d k E(k) = < v_ {\alpha }v_{\alpha} > $ and
$ \int d k P(k) = < \dot{\omega} \dot{\omega} > $ is the power spectrum
associated with $\dot{\omega} $. Numerical simulations of 2d
turbulence are in broad agreement with $Q(k) $ having slope $-3$ [9].
In the present context we can also search for the generalizations (if any)
of such scaling laws, which might also provide a useful check in any
future simulations in $2d$ turbulent MHD and its dyonic counterpart.

Let us consider the two limiting cases  (I) $\cB = V = 0$ and (II)
$ \ccA =V=0$ for which we can unambiguously associate a conformal field
to the physical  electric and magnetic fields as discussed earlier.
In case (I) the only conserved quadratic invariants are $\int d k
{\cal K} $  and $\int d  k \ccA \ccA $. Defining $\int d k P_{A}  =
< \dot \ccA \dot \ccA > $ and $ Q^{I} = ( {\cal K} P_A k^{-8} {)}^{1/4} $
one finds for constant $\ccA$ flux solutions to the Hopf equations
(\ref{eq:5.4} ) that $Q^{I} \sim k^{-2} + k^{\Delta_\psi + \Delta_\chi} $. The
$-2$ contribution to the slope of $Q^{I} $
is universal but there are also non universal terms which  may,
depending on the values of $\Delta_\psi $ and $\Delta_\chi $ be
small compared to this.  In case (II) there are three conserved
quadratic invariants $\int d k E(k) ,  < \omega \omega > $ and
$ \int d k E_{\ccB} = < \ccB \ccB > $.  In this case apart from the
universal scaling law for $Q(k) $ defined above
, there is a new
scaling law for the quantity
$Q^{II}  = ( E_{\ccB} P_{\ccB} k^{-6} {)}^{1/4 } $
where $\int d k P_{\ccB} = < \dot{\ccB} \dot{\ccB} > $ . In the inertial
range with constant $\ccB $ flux  defined earlier, one finds
$Q^{II} (k)  \sim k^{-2} $.

To summarize, we have analysed the conformal field theoretic solutions of
various inertial ranges of $2d$ duality invariant Magnetohydrodynamics.
We focused on the equations of dyonic plasmas in the ideal limit. These are
equivalent to those of normal ideal MHD equations but with the physical
magnetic and electric fields replaced by the duality invariant quantities
$\cB$ and $\cE.$ Due to the extra structure of the system coming from
electric-magnetic duality symmetry, new conditions have to be imposed on the
OPE's  of CFT solutions. These conditions arise in the attempt of associating
conformal fields to the physical variables.
This is essential in determining the contribution of the magnetic
and electric fields to the power spectrum of the system.
\vskip0.5in
\begin{center}
{\bf\Large Acknowledgements}
\end{center}
\vskip0.1in
The work of S.Thomas was supported by the Royal Society of Great Britain
and that of W. A. Sabra  by P. P. A. R. C.

\begin{center}
\noin{\bf\Large References}
\end{center}
\begin{description}
\item{[1]} O. Coceal, W. A. Sabra and S. Thomas, {\it Duality Invariant
Magnetohydrodynamics And Dyons}, preprint QMW-PH-96-02, hep-th/9603200.
\item{[2]} O. Coceal and S. Thomas, {\it Conformal Models of
Magnetohydrodynamic
Turbulence}, preprint QMW-PH-95-45, hep-th/9512022.
\item{[3]} D. R. Nicholson, {\it Introduction to Plasma Theory}, Wiley (New
York) 1983.
\item{[4]} P. Goddard and D. I. Olive, {\it Rep. Prog. Phys.} {\bf41} (1978)
1357.
\item{[5]} G. Ferretti and Z. Yang, {\it Europhys. Lett.} {\bf 22}  (1993) 639.
\item{[6]}  P. Olesen, {\it Phys. Lett.} {\bf B366} (1996) 117.
\item{[7]} A. M. Polyakov {\it Nucl. Phys.} {\bf B396} (1993) 367.
\item{[8]} D. A. Lowe {\it Mod. Phys. Lett.} {\bf A8} (1993) 923.
\item{[9]} R. Benzi, B. Legras, G. Parisi and R. Scardovelli,  {\it  Europhys.
Lett.,} {\bf 29} (1995) 203.
\item{[10]} M. R. Rahimi Tabar and S. Rouhani,  {\it Ann. Phys.} {\bf 246}
(1996) 446.
\item{[11]} A. A. Belavin, A. M. Polyakov and A. B. Zamolodchikov,
{\it Nucl. Phys.} {\bf B241} (1984) 333.
\end{description}
\end{document}